\def\Murcia{Departamento de Matem\'atica Aplicada, Facultad de Inform\'atica, Campus
de Espinardo, 30100 Murcia, Spain}

\def\IAA{Instituto de Astrof\'{\i}sica de Andaluc\'{\i}a, Apartado Postal 3004,
18080 Granada, Spain}

\def\CarlosI{Instituto de F\'\i sica Te\'orica y Computacional Carlos I,
Facultad de Ciencias, Universidad de Granada, Campus de Fuentenueva, 
Granada 18002, Spain} 

\def\Comision{Work partially supported by the DGICYT.}

\def\bdm{\begin{displaymath}} 
\def\edm{\end{displaymath}} 
 
\def\bea{\begin{eqnarray}} 
\def\eea{\end{eqnarray}} 
 
\def\be{\begin{equation}} 
\def\ee{\end{equation}} 
 
\def\ba{\begin{array}} 
\def\ea{\end{array}}

\def\Gt{$\widetilde{G}\;$}
\def\Gtm{\widetilde{G}\,}

\def\Gtc{$\check{G}\;$}

\def\calG{ {\cal G} }

\def\R{{{\mathbb R} }} 

\def\calG{{\cal G}}

\def\ni{\noindent}
\def\nn{\nonumber}

\def\w{\omega}

\newcommand{\XL}[1]{ {\tilde{X}}^{L}_{#1} }

\newcommand{\XR}[1]{ {\tilde{X}}^{R}_{#1} }

\newcommand{\XLc}[1]{ {\check{X}}^{L}_{#1} }

\documentclass[12pt]{article} 
\usepackage{amsfonts,amssymb} 
 
\begin{document}


\begin{center}
{\LARGE {\bf Group-cohomology refinement to classify $G$-symplectic manifolds$^1$ }} 
\end{center}

\vskip 0.2 cm

\centerline{J. Guerrero$^{2,4}$, J.L. Jaramillo$^{2,3}$ 
       and  V. Aldaya$^{2,3}$}

\footnotetext[1]{\Comision} \footnotetext[2]{\IAA} 
\footnotetext[3]{\CarlosI} \footnotetext[4]{\Murcia}

 \begin{center}
{\bf Abstract} 
\end{center}

\small

 \begin{list}{}{\setlength{\leftmargin}{3pc}\setlength{\rightmargin}{3pc}}
\item ``Pseudo-cohomology'', as a refinement of Lie group cohomology, is soundly studied 
aiming at  classifying of the symplectic manifolds associated with Lie groups.
In this study, the framework of symplectic cohomology provides fundamental new insight, 
which enriches the analysis previously developed in the setting of Cartan-Eilenberg 
$H^2(G,U(1))$ cohomology.

\end{list}

\normalsize 



%
%
%
%

%
%
%
%
%
%

\section{Introduction}

{}From the strict mathematical point of view, the orbits of the coadjoint 
representation of Lie groups provide a source for symplectic 
manifolds on which a given Lie group acts as a group of symplectomorphisms,
i.e. $G$-symplectic manifolds. Even, for finite-dimensional semi-simple groups, 
this mechanism essentially exhausts all models of they. These $G$-symplectic
manifolds could then be considered as phase spaces of physical systems for
which $G$ can be called the ``basic symmetry''. From the physical point of view,
however, the simplest physical systems (the free non-relativistic particle,
for instance) possess a phase space endowed with a symplectic form
whose associated Poisson bracket realizes the Lie algebra of a central
extension  of
the basic ``classical'' symmetry. Central extensions of Lie groups by $U(1)$, 
associated with projective unitary representations, were classified long ago by 
Bargmann \cite{Bargmann} by means of the cohomology group $H^2(G,U(1))$
\cite{C-E}. Later, the momentum map from the phase space to
the coalgebra ${\cal G}^*$ of the basic ``classical'' symmetry group,
constructed with the set of Noether invariants of the physical system,
was used by Souriau \cite{So70} to define the symplectic cohomology group
$H^1_S(G,{\cal G}^*)$ characterizing equivalently the central extensions of
a simply-connected group $G$.

In this paper we revisit the notion of Lie group ``pseudo-cohomology'' in an
attempt to classify all possible (quantizable) $G$-symplectic manifolds for
an arbitrary Lie group $G$, in such a way that both coadjoint orbits and phase
spaces realizing central extensions can be put together into (``pseudo''-)cohomology
classes. By the way, the prefix ``pseudo'' had its origin \cite{AA85} in the fact
that the corresponding central extensions are trivial from the mathematical
point of view, the associated cocycle being a coboundary, although they behave as
if they were non-trivial in some aspects, as we shall show.

Our study here is made in the language of symplectic
cohomology of Lie groups. The insight provided by the natural and explicit
role of phase spaces in symplectic cohomology offers a
more intuitive understanding of the significance of pseudo-cohomology in
classifying dynamics, as well as  an easier mathematical
handling which allows a generalization of the mathematical results
obtained in its original presentation \cite{Marmo,medida}.

Roughly speaking, pseudo-cohomology emerges as a refinement of the
equivalence classes of 2-cocycles in the cohomology group
$H^2(G,U(1))$. The first clues for the need of such a refinement occurred
when studying the problem of the In\"on\"u-Wigner contraction of centrally
extended Lie groups (see Saletan \cite{Saletan}). An example of this need
appears in the contraction Poincar\'e $\rightarrow$ Galileo where a
special kind of trivial 2-cocycles in the Poincar\'e group, becomes true
2-cocycles for the Galileo group in the $c\rightarrow\infty$ limit.
The underlying reason is
that, while the 2-cocycle is well-behaved in the limit, its generating
function is not, thus occurring a generation of cohomology \cite{AA85}.
The second indication for the need of pseudo-cohomology appeared in the
context of generalized Hopf fibrations of semi-simple Lie groups related
to \v{C}ech (true, i.e. non-coboundary) cocycles of coadjoint orbits. 
Such pseudo-cocycles play in fact
a fundamental role in the explicit construction of the local exponent
associated with Lie-algebra cocycles of the corresponding Kac-Moody groups
\cite{AN87}.


In spite of these antecedents, the importance of
pseudo-cohomology is more evident in the framework of Group Approach to
Quantization (GAQ), a group theoretical quantization scheme designed for
obtaining the dynamics of a physical system out of a Lie group
\cite{JMP23}. In particular, GAQ starts from a central extension \Gt of a
Lie group $G$ by $U(1)$ in such a way that the symplectic form of the
classical phase space is derived from the 2-cocycle which defines the
central extension. Nevertheless, the correspondence between central
extensions and symplectic forms is not one-to-one. The most obvious
illustration of this is the case of groups with trivial cohomology group
$H^2(G,U(1))$ (such as the Poincar\'e group in 3+1 dimensions or
finite-dimensional semi-simple groups). In fact, even though these
groups do not admit
non-trivial central extensions, genuine symplectic structures and dynamics
can be derived out of them \cite{AA85,AN87}.
 The rationale for this is the existence of 2-cocycles which are
coboundaries, and therefore trivial from
 the cohomological point of view,
 but which do define authentic symplectic structures.  Coboundaries with
this property are called {\it pseudo-cocycles}, giving rise to trivial
central extension referred to as {\it pseudo-extensions}.

The study of this mechanism and the characterization of the classes of
pseudo-extensions associated with non-equivalent symplectic structures,
led in an explicit way to the notion of {\bf pseudo-cohomology},
constituting the more systematic and clarifying approach to the problem
\cite{Marmo,medida}. This {\it standard} view of pseudo-cohomology is
described in the next section.

\section{Pseudo-cohomology in GAQ}
\label{pseudocohomologia}

As commented in the introduction, GAQ is a formalism devised for obtaining
the (quantum or classical) dynamics of a physical system out of a Lie
group (of its symmetries). The starting point is a central extension \Gt
of the symmetry group $G$ by $U(1)$ (or $\mathbb{R}$ to recover the
classical dynamics), determined by a 2-cocycle (local exponent) $\xi:
G\times G\rightarrow \mathbb{R}$. The group law then reads:
\begin{equation}
g\:''=g\:'*g\,,\qquad \zeta\:'' = \zeta\:' \zeta e^{i\xi(g\:',g)}
\end{equation}

\ni where $g\:'',g\:',g\in G$ and $\zeta\:'',\zeta\:',\zeta \in U(1)$. On
the Lie group \Gt we have at our disposal left and right-invariant vector
fields. If we choose a coordinate system $(\{g^i\}_{i=1}^{\rm dim
G},\zeta)$ in \Gt, a basis for the vector fields is
given by $\XL{i}$ and $\XR{i}$, respectively, and their dual sets of left
and right invariant 1-forms are denoted by $\theta^{L(i)}$ and
$\theta^{R(i)}$, respectively. One of the left-invariant 1-forms,
$\Theta\equiv \theta^{L(\zeta)}$, the $U(1)$-component of the
left-invariant canonical 1-form on the Lie group \Gt, is chosen as the
connection 1-form of the principal bundle $U(1)\rightarrow \Gtm
\rightarrow G$, thus defining a notion of {\it horizontality}.

This connection 1-form, referred to as quantization 1-form, depends directly on the
2-cocycle $\xi$ and can be
used to define a symplectic structure in a unique manner. In fact, if
${\cal G}_\Theta$ is the characteristic distribution of $\Theta$, i.e. the
intersection of the kernel of $\Theta$ and $d\Theta$, then $\Gtm/{\cal
G}_\Theta$ is a quantum manifold $P$ \cite{JMP23,Woodhouse}. This means
that $\Gtm/{\cal G}_\Theta$ is a contact manifold with contact 1-form
$\Theta|_P$. The Quantum manifold $P$ is in turn a $U(1)$
Principal bundle $U(1)\rightarrow P \stackrel{\pi}{\rightarrow} S$ with
base a symplectic manifold, $S=P/U(1)$ endowed with a symplectic form
$\omega$ such that $\pi^*\omega =d\Theta$.

The symplectic structure $(S,\omega)$ is not completely determined by the
cohomology class to which the 2-cocycle $\xi$ belongs. In fact, different
yet cohomologous 2-cocycles can lead to completely different symplectic
structures $(S,\omega)$ (think, for instance, of a semi-simple Lie group,
with trivial cohomology but with many different kinds of symplectic
structures determined by its coadjoit orbits). This phenomenon suggests,
again, a refinement in the classification of 2-cocycles in such a way that
a one-to-one correspondence between the refined classes and the symplectic
structures could be established. This refinement will define
pseudo-cohomology. Therefore, the latter is intrinsically tied to the
classification of possible symplectic structures constructed out of a Lie 
group. 

The main idea for the definition of these subclasses in $H^2(G,U(1))$ can 
be intuited from the expression of $\Theta$ in terms of the 2-cocycle 
$\xi$: 
\begin{equation}
\Theta = \frac{d\zeta}{i\zeta} + \left.\frac{\partial \xi(g',g)}{\partial 
g^i}\right|_{g'=g^{-1}} dg^i \label{Theta} 
\end{equation}

If, now, a 2-coboundary $\xi_\lambda (g',g) 
=\lambda(g'*g)-\lambda(g')-\lambda(g)$ generated by the function 
$\lambda:G\rightarrow \mathbb{R}$, is added to $\xi$, the expression for 
the new quantization 1-form $\Theta'$ (as the $U(1)$-component of the 
canonical 1-form for the centrally extended Lie group defined by 
$\xi+\xi_\lambda$) is given by: 
\begin{equation}
\Theta' = \Theta + \Theta_{\lambda}=\Theta+ \lambda^0_i\theta^{L(i)} 
-d\lambda \label{Thetaprima} 
\end{equation}

\ni where $\lambda^0_i\equiv \frac{\partial \lambda(g)}{\partial 
g^i}|_{g=e}$. Thus, the new term $\Theta_{\lambda}$ added to the 
connection 1-form $\Theta$ by the inclusion of a 2-coboundary depends
only, up to a total differential, on the gradient at the identity
$\lambda^{0}$ of the generating function $\lambda(g)$. In fact, if we 
denote $\Theta_{\lambda^0}=\lambda^0_i\theta^{L(i)}$, then the total 
differential disappears when the presymplectic 2-form $d\Theta$ is 
considered, in such a way that $d\Theta'=d\Theta + d\Theta_{\lambda^0}$.

{}From these considerations two conclusions can be drawn: 
\begin{itemize}
\item[{\it i)}] A 2-coboundary contribute non-trivially to
the connection 1-form $\Theta$ and to the symplectic structure determined 
by $d\Theta$, if and only if $\lambda^0\neq 0$. 

\item[{\it ii)}] This contribution depends only (up to a total differential, 
which does not affect the symplectic structure) on the local properties of 
the generating function $\lambda(g)$ at the identity of the group, through 
its gradient at the identity $\lambda^0$. 
\end{itemize}

A 2-coboundary $\xi_\lambda$ such that $\lambda^0\neq 0$ is named a {\it 
pseudo-cocycle}. The name reflects the fact that they are trivial 
2-cocycles but, from the dynamical point of view, behave as if they were 
non-trivial. If we consider the group $G$ centrally extended by this 
pseudo-cocycle $\xi_\lambda$, the extended group \Gt is isomorphic to 
$G\times U(1)$. However, we will refer to this extension as a {\it 
pseudo-extension}, to underline the fact that, although trivial
as a central extension, it can lead to a non-trivial symplectic structure and 
non-trivial dynamics. 

\medskip

The next point to explore is the conditions under which two different 
2-coboundaries, $\xi_\lambda$ and $\xi_{\lambda'}$, generated by functions 
$\lambda$ and $\lambda'$ with different gradients at the identity 
$\lambda^0$ and $\lambda'{}^0$, determine the same symplectic structure 
$(S,\omega)$, up to symplectomorphisms. This condition will define a 
refined equivalence relation inside each cohomology class. 
For the sake of simplicity, we shall restrict ourselves to simply
connected Lie-groups.

The clue in the definition of the new equivalence relation is given by the 
fact that $\lambda^0$ defines an element of ${\cal G}^*$, the dual of the 
Lie algebra ${\cal G}$ of $G$, usually named the {\it coalgebra}. This 
can be seen by noting that $\Theta_{\lambda^0}=\lambda^0_i\theta^{L(i)}$ 
defines, at the identity of $G$, an element of ${\cal G}^*$ given by 
$\Theta_{\lambda^0}|_{g=e}=\lambda^0$. It is also important to note that 
$\Theta_\lambda|_{g=e} = 0\in {\cal G}^*$ (due to the presence of 
$d\lambda$), in such a way that the quantization 1-form $\Theta$ verifies 
$\Theta|_{g=e}=( 0,\ldots,0,1)\in \tilde{\cal G}^*$, whatever the 
2-cocycle $\xi$ we are considering (here $\tilde{\cal G}^*$ is the dual of 
the extended algebra $\tilde{\cal G}$ associated with the extended group 
\Gt). This fact will be of relevance in the relationship between 
pseudo-cohomology and symplectic cohomology. 

Once we have established that $\lambda^0 \in {\cal G}^*$, it is natural to 
propose their classification in accordance with  the coadjoint orbits. 
This will prove to be the 
correct ansatz, provided we use the correct coadjoint action. 

A bit of notation is in order. Let us denote the equivalence class of the 
cocycle $\xi$, defining a certain central extension \Gt of $G$ by $U(1)$, 
by $[[\xi]]\in H^2(G,U(1))$. We are going to introduce a further partition 
in each class $[[\xi]]$ into equivalence subclasses $[\xi]$. 

For the sake of clarity, we 
 shall firstly define this partition
for the trivial cohomology class $[[\xi]]_0$, made out of trivial 
cocycles, i.e. 2-coboundaries $\xi_\lambda$. This would be enough for 
groups with trivial cohomology $H^2(G,U(1))=0$ (that is, with only the 
trivial class), such as finite-dimensional semi-simple groups or the 
Poincar\'e group (in 
$3+1$ dimensions). It is also valid for fully centrally-extended groups 
\Gt, for which $H^2(\Gtm,U(1))=0$. The case of groups with non-trivial 
cohomology or non-fully  central-extended groups $\tilde{G}$, with 
$H^2(\Gtm,U(1))\neq 0$, will be considered in section \ref{notrivial}.



 \subsection{The trivial class}
 \label{trivial}



Given a Lie group $G$, a natural action of $G$ on ${\cal G}^{*}$ is 
provided by the coadjoint action $Coad$, defined as the dual of the 
adjoint action of $G$ on ${\cal G}$. More explicitly, with the adjoint 
action of $G$ on ${\cal G}$ given by $Ad 
\;g(X)=(R^T_{g^{-1}}L^T_g)(e)\cdot X$, where $g\in G$ and $X\in {\cal G}$ 
\footnote{Throughout the paper, the differential of a given application 
will be denoted with a superscript T.}, the coadjoint action $Coad:
G\rightarrow Aut({\cal G}^*)$ has the form $Coad (g)\mu(X)=\mu(Ad \;
g^{-1}(X))$, where $\mu \in {\cal G}^*$. It is also convenient to make 
explicit the infinitesimal version of this action. Linearizing on the $g$ 
variable we obtain the coadjoint action of the Lie algebra on the 
coalgebra: $(Coad)^T(e)\equiv coad:{\cal G}\rightarrow End({\cal G}^*)$. 
Its explicit expression is given by $coad\; 
X(\mu)(Y)=\mu(ad\;X(Y))=\mu([X,Y])$, with $X, Y\in {\cal G}$ and 
$\mu\in{\cal G}^*$.

The orbits of this action are specially relevant in our study. Given a 
point $\mu\in{\cal G}^{*}$, the orbit through this point by the action of
the whole group $G$ is $Orb(\mu)=\{Coad(g)\mu\,\,/\,\, g\in G\}$,
diffeomorphic to 
$G/{G_\mu}$ where $G_\mu$ is the isotropy group of $\mu$. The coadjoint 
action determines a foliation of ${\cal G}^{*}$ in orbits, in such a way 
that any point belongs to one (and just one) orbit (by the definition, the 
point $\mu$ belongs to $Orb(\mu)$), and two points in the same orbit are 
always connected by the coadjoint action. 

Coadjoint orbits of Lie groups are interesting from the physical point of 
view since they possess a natural symplectic structure $(Orb(\mu),\omega)$ 
with the symplectic form given by 
\bea 
\omega_\nu(X_\nu,Y_\nu)=\nu([X,Y]), 
X_\nu,Y_\nu \in T_\nu (Orb(\mu)) \ \ , \label{symplcoad} \eea 
where $\nu\in Orb(\mu)\subset{\cal G}^*$, $X_{\nu},Y_{\nu} \in T_\nu (Orb(\mu))$ and
$X\in {\cal G}$ is related to $X_\nu\in T_\nu Orb(\mu)$ by $X_{\nu}=coad(X)
\;\nu$, and analogously for $Y_\nu$ and $Y$ (note we are using the 
fact that ${\cal G}^*$ is a linear space in order to identify its points 
with tangent vectors).

\medskip
There is a close relationship between pseudo-extensions and coadjoint 
orbits, that can be stated as follows. A pseudo-extension characterized by
the generating function $\lambda(g)$ with gradient at the identity 
$\lambda^0\neq 0$ defines a presymplectic form 
$d\Theta_{\lambda}=d\Theta_{\lambda^0}$ depending only on $\lambda^0$. In
the trivial case we are discussing in this section, the quotient of \Gt by 
the characteristic subalgebra ${\cal G}_ {\Theta_{\lambda^0}} \equiv \ker 
\Theta_\lambda \cap \ker d\Theta_{\lambda^0}$, defines a quantum manifold
$P$, and the quotient $S=\Gtm/(G_{\Theta_{\lambda^0}}\times U(1))\sim
G/G_{\Theta_{\lambda^0}}$ is a symplectic manifold with symplectic form
$\omega_{\lambda^0}$ given by $\pi^*\omega_{\lambda^0}=
d\Theta_{\lambda^0}$, where $\pi:\,P\rightarrow S$ is the canonical
projection and $G_{\Theta_{\lambda^0}}$ is the (connected) subgroup associated with
${\cal G}_ {\Theta_{\lambda^0}}$. $G/G_{\Theta_{\lambda^0}}$ is in fact locally diffeomorphic to a
coadjoint orbit (the one passing through $\lambda^0$) . This can be seen by noting:
\begin{itemize}
\item[a)] The pre-symplectic form adopts the expression:
\bea
d\Theta_{\lambda^0}=\frac{1}{2}\lambda^0_kC^k_{ij} \theta^{L(i)}\wedge \theta^{L(j)} \ \ ,
\eea
when using the Maurer-Cartan equations, and
therefore,
\bea
d\Theta_{\lambda^0}(X^L_i,X^L_j)=\lambda^0_kC^k_{ij}=\lambda^0([X^L_i,X^L_j])
\eea
where $\{X^L_i\}$ is a basis for ${\cal G}$ and $\lambda^0\in{\cal
G}^*$, thus reproducing (\ref{symplcoad}) (before falling down to the
quotient).
\item[b)] The characteristic group $G_{\Theta_{\lambda^0}}$ coincides with (the connected component of)
the isotropy group of $\lambda^0$, $G_{\lambda^0}$ under the coadjoint action,
thus defining (locally) the same quotient space. At the infinitesimal level, a
vector $Y=Y^iX^L_i$ belongs to ${\cal G}_{\Theta_{\lambda^0}}$ iff
$Y^i\lambda^0_kC^k_{ij}=0, \forall j$, which is the same condition for $Y$
to belong to ${\cal G}_{\lambda^0}$.
\end{itemize}


Using the transformation properties of left-invariant one-forms under
translation by the group, it is easy to check that:
\begin{equation}
Ad(g)^{*} (\Theta_{\lambda^{0}})= \Theta_{Coad(g)\lambda^{0}}
\end{equation}

\ni where $Ad(g)^{*}$ denotes the pull-back of the adjoint action of the
group on itself (conjugation), acting on $\theta^{L(i)}$, and on the right
hand side $Coad(g)$ acts on $\lambda^{0}$.

Although the connection 1-form is given by $\Theta_\lambda$ rather than
$\Theta_{\lambda^0}$, the symplectic form is determined by just
$\lambda^0$, and it transforms in a similar way:
\begin{equation}
Ad(g)^{*} (d\Theta_{\lambda^{0}})= d\Theta_{Coad(g)\lambda^{0}}
\end{equation}

These results can be summarized in the following proposition:

\medskip

\ni{\bf Proposition 1.} {\it Let $G$ be a Lie group and consider two
coboundaries $\xi_{\lambda_1}$ and $\xi_{\lambda_2}$ with generating
functions $\lambda_1(g)$ and $\lambda_2(g)$, defining the (trivial)
central extensions $\tilde{G}_1$ and $\tilde{G}_2$, respectively. If 
$\Theta_{\lambda_1}$ and $\Theta_{\lambda_2}$ are the Quantization one-forms associated with each group, 
with $G_{\Theta_{\lambda_1^0}}$ and $G_{\Theta_{\lambda_2^0}}$ as their respective characteristic
subgroups, the two symplectic spaces $\tilde{G}_1/ (G_{\Theta_{\lambda_1^0}}\times U(1))$ and
$\tilde{G}_2/(G_{\Theta_{\lambda_2^0}}\times U(1))$, with symplectic forms given by $\omega_{\lambda_1^0}$ and
$\omega_{\lambda_2^0}$ such that $d\Theta_{\lambda_1^0}=\pi^*\omega_{\lambda_1^0}$ and
$d\Theta_{\lambda_2^0}=\pi^*\omega_{\lambda_2^0}$ respectively,
are symplectomorphic if there exists $h\in G$ such that:
\bea 
\lambda_1^0=Coad(h)\lambda_2^0, \eea the symplectomorphism being 
given by $Ad(h)$: 
\bea 
d\Theta_{\lambda_1^0}=Ad(h)^*d\Theta_{\lambda_2^0}
\eea }

\ni {\it Proof:} It simply remains to proof that the two spaces $\tilde{G}_1/ (G_{\Theta_{\lambda_1^0}}\times U(1))$
and $\tilde{G}_2/(G_{\Theta_{\lambda_2^0}}\times U(1))$ are diffeomorphic.  Since the extensions are trivial,  
$\tilde{G}_i$, $i=1,2$ are isomorphic to $G\times U(1)$, therefore $\tilde{G}_i/ (G_{\Theta_{\lambda_i^0}}\times 
U(1))\approx G/G_{\Theta_{\lambda_i^0}}$, $i=1,2$. If $\lambda_1^0=Coad(h)\lambda_2^0$, then $G_{\Theta_{\lambda_1^0}}$ and 
$G_{\Theta_{\lambda_2^0}}$ are conjugated subgroups by the adjoint action and this implies that the
two spaces $G/G_{\Theta_{\lambda_i^0}}$, $i=1,2$ are diffeomorphic.

\ni This suggests us to define the equivalence relation in $[[\xi]]_0$ in 
the following way: 
 
\medskip

\noindent {\bf Definition 1:}\,\, {\it Two coboundaries $\xi_\lambda$ and
$\xi_{\lambda'}$ with generating functions 
 $\lambda$ and $\lambda'$, respectively, belong to the same equivalence
subclass $[\xi]$ of $[[\xi]]_0$ if and only if the gradients at the 
identity of the generating functions are related by:} 
\be
\lambda^0{}'= Coad(g)\lambda^0\,, \ee 

{\it \ni for some $g\in G$}. 

\medskip
 
We shall denote by $[\xi]_{\lambda^0}$ the equivalence class of 
coboundaries ``passing through'' $\lambda^0$. The equivalence relation 
introduced in this way will be named {\bf pseudo-cohomology}\footnote{This 
equivalence relation does not define, in general, a cohomology.}, and
the subclass of (trivial) central extensions 
defined by all $\xi_\lambda\in [\xi]_{\lambda^0}$ will be called the 
central {\bf pseudo-extension} associated with $[\xi]_{\lambda^0}$. 

The condition $\lambda^0{}'= Coad(g)\lambda^0$ means that $\lambda^0$ and 
$\lambda^0{}'$ are related by the coadjoint action of the group $G$. 
Therefore, $\lambda^0$ and $\lambda^0{}'$ lie in the same coadjoint orbit 
of $G$ in $\calG^*$. A pseudo-cohomology class is therefore directly 
associated with a coadjoint orbit in ${\cal G}^*$. 
 
If $\lambda^0{}'=\lambda^0$, then $\xi_\lambda$, $\xi_{\lambda'}\in 
[\xi]_{\lambda^0}$. Thus, we can always choose a representative element in 
each subclass "linear" in the local coordinate system, 
$\xi_{\lambda^0}(g)=\lambda^0_i g^i$. If the local coordinates $\{g^i\}$ 
are canonical, and if we restrict ourselves to canonical 2-cocycles (see 
\cite{Bargmann}), then two cohomologous 2-cocycles differ in a 
2-coboundary $\xi_\lambda$ with $\lambda(g)$ linear in the canonical 
coordinates. Then pseudo-cohomology is a further partition of "linear" 
coboundaries into equivalence classes through the coadjoint action of the 
group $G$ on ${\cal G}^*$ (for the trivial class, at the moment).

However, the correspondence between pseudo-cohomology classes and
coadjoint orbits for a Lie group $G$ is not onto. The relation is
established in the following theorem:

\medskip

\noindent {\bf Proposition 2:}\,\, {\it Pseudo-cohomology classes are associated
with coadjoint orbits which satisfy an integrality condition: the
symplectic 2-form $\w$ naturally defined on the coadjoint orbit by
(\ref{symplcoad}) has to be
of integer class. }
\medskip

In fact, this integrality condition is required for $\xi_{\lambda^0}$ to
define a global coboundary on $G$;
non-integral coadjoint orbits of $G$ cannot be related to central
pseudo-extensions of $G$, since they do not define a proper (global) Lie
group. \\
{\it Proof:} A (pseudo-) centrally extended Lie group gives rise to a Quantum manifold in
the sense of Geometric Quantization (see Sec. 2) when taking quotient by the characteristic subalgebra
\cite{JMP23}. Therefore, as a consequence of the necessary and sufficient condition for the existence of a
quantization of a given symplectic manifold (see \cite{Woodhouse,JMP23}), the closed 2-form on the coadjoint
orbit is of integer class.


Let us see another way of looking at the integrality condition. The vector
$\lambda^0$ is an element of $\calG^*$ and, therefore, it is a linear
mapping from $\calG$ to $\R$. It is easy to check that when restricted to
$\calG_{\lambda^0}$ (the Lie algebra of the isotropy group $G_{\lambda^0}$
of the coadjoint orbit passing through $\lambda^0$), $\lambda^0$ defines one
dimensional representation of the latter. Then, the integrality condition
on the coadjoint orbit parallels the requirement for $\lambda^0$ of being 
exponentiable (integrable) to a unitary character of the group 
$G_{\lambda^0}$ (note however that this remark resorts to the level of 
representation theory of the group, whereas the above-stated theorem involves
only the Lie group structure). 
 
This relationship between integrality condition of the coadjoint orbit and 
"integrality" of the character defined by $\lambda^0$ reveals, in passing,
that the 
Coadjoint Orbit Method of Kostant-Kirillov \cite{Kostant,Kirillov}, 
intended to obtain unitary irreducible representations of Lie groups
using (what in Physics is now known as) Geometric Quantisation
\cite{Woodhouse} on coadjoint orbits of Lie groups, is a particular case 
of the induced representation technique of Mackey \cite{Mackey}. 
 
Let us denote by \Gtc the central pseudo-extension of $G$, characterised 
by $\xi_{\lambda^0}$. It defines a central pseudo-extension 
of $\calG$: 
\be [\XLc{i},\XLc{j}]=C_{ij}^k(\XLc{k} + \lambda^0_k X_0)\,, \ee 
 
\ni where $X_0$ is the (central) generator associated with $U(1)$ (our 
convention is to take $X_0=i I$ in any faithful unirrep of \Gtc). The 
left-invariant vector fields of \Gtc (denoted with check) are related to 
those of $G$ by: 
\be \XLc{i}= X^L_i + (X^L_i\lambda-\lambda^0_i)X_0\,, \ee 

\ni with a similar relation for the right-invariant vector fields.
 
{}From this point of view, central pseudo-extensions are on the same 
footing as true (non-trivial) central extensions, and we can employ with 
them the same techniques for obtaining (projective) unirreps of $G$ 
(specially for semisimple Lie groups). Once a projective representation of 
$G$ (which is a true representation of \Gtc, the pseudo-extended group) 
has been obtained in this way, in order to obtain the true 
(non-projective) representations of $G$ associated with it we simply 
redefine the generators in the following way\footnote{Or, equivalently, 
the functions of the Hilbert space carring the representation should be 
redefined multiplying them by an appropriate factor.}
%
\be \XLc{i} \rightarrow \XLc{i}+\lambda^0_i X_0 = X^L_i + 
(X^L_i\lambda)X_0\,. \ee 

\subsection{Non trivial classes}
\label{notrivial}

Let us consider now non-trivial cohomology classes $[[\xi]]\neq 
[[\xi]]_0$, in the case of groups $G$ with non-trivial cohomology, or 
extended groups \Gt which still admit further central extensions, that is, 
with $H^2(\Gtm,U(1))\neq 0$. 

In order to proceed, a representative element $\xi\in [[\xi]]$ must be 
chosen. 
%
%
We can add to $\xi$ a coboundary $\xi_{\lambda}$ generated by a function 
$\lambda$, with non-trivial gradient at the identity of $G$. The resulting
cocycle $\xi'=\xi+\xi_\lambda$ defines a new central extension $\Gtm'$ of 
$G$ isomorphic, from the group-theoretical point of view, to \Gt. The 
question is whether these pseudo-extensions can be classified into 
equivalence classes leading to the same symplectic structures, as in the 
case of the trivial class of Sec. \ref{trivial}. The na\"\i ve 
classification in coadjoint orbits of the group $G$ does not work in this 
case (since there is a mixture of true cohomology and pseudo-cohomology), 
and there is no clue, at this level, of how the classification should be 
done. 

The direct relation between pseudo-cohomology and coadjoint orbits 
obtained for the trivial class, allows us to resort to symplectic 
cohomology, as a tool for classifying symplectic structures (see Sec. 
\ref{symplectic}), to come in our help. In this framework, it will be 
shown that a classification of pseudo-cocyles in the non-trivial classes 
is possible and entails a slight generalization with respect to that of 
the trivial class, in the sense that the classification should be done
using the deformed coadjoint action (associated with the central extension 
determined by the non-trivial 
 class we are 
considering).


\section{Symplectic Cohomology}
\label{symplectic} In the previous section we have seen how GAQ can be 
used to define symplectic structures out of a Lie group, naturally leading 
to the notion of pseudo-cohomology. In this section we review a different 
approach to the discussion of the symplectic structures defined in terms
of a Lie group $G$. Firstly, we briefly recall the fundamentals of the
so-called symplectic cohomology of $G$. The rationale for this structure
can be found in the context of momentum mapping (\cite{So70} and below).
Secondly, we use this mathematical structure to classify a family of
symplectic spaces which generalize the ones obtained by the coadjoint
action of a group $G$.

\subsection{Lie group cohomology. Symplectic cohomology}
Given a Lie group $G$, an abelian Lie group $A$ and a (left) action $L$ of
$G$ on $A$, we define the n-cochains $\gamma_n$ as mappings \bea
\gamma_n:G\times\ldots^{n)}\times G\rightarrow A \ \ , \eea in such a way
that the standard sum of mappings \bea
(\gamma_n+\gamma'_n)(g_1,\ldots,g_n)=\gamma_n(g_1,\ldots,g_n) +
\gamma'_n(g_1,\ldots,g_n) \ \ . \eea endows the space of n-cochains,
denoted as $C^n_L(G,A)$, with the structure of an abelian group.

\ni The coboundary operators $\delta:C^n_L(G,A)\rightarrow C^{n+1}_L(G,A)$ are
defined by \bea (\delta\gamma_n)(g_1,\ldots,g_n,g_{n+1})&\equiv&L(g_1)
(\gamma_n)(g_2,\ldots,g_n,g_{n+1})+ \nn\\
&+&\sum_{i=1}^{n}(-1)^i\gamma_n(g_1,\ldots,g_ig_{i+1},g_{i+2},
\ldots,g_{n+1})+ \label{a1}\\ &+&(-1)^{n+1}\gamma_n(g_1,\ldots,g_n) \ \ ,
\nn \eea satisfying the nilpotency condition
$\delta\circ\delta=0$.
We can define the subspaces of $n-$cochains
$Z^n\equiv Ker(\delta)\subset C^n_L(G,A)$, whose elements are {\it closed
n-cochains} and
$B^n \equiv Im(\delta)\subset C^n_L(G,A)$, whose elements are {n-coboundaries}.
Two exact n-cochains are equivalent if their difference is a
coboundary. Cohomology groups are defined by this equivalence
\bea
H^n_L(G,A)=\frac{Z^n}{B^n} \ \ \eea
and their elements are called {\it n-cocycles}.
For the first cohomology groups the
expression of (\ref{a1}) takes the form,
\bea
(\delta\gamma_0)(g)&=&L(g)\gamma_0-\gamma_0 \nn \\
(\delta\gamma_1)(g_1,g_2)&=&L(g_1)\gamma_1(g_2)-\gamma_1(g_1g_2)+
\gamma_1(g_1) \label{a2} \\
(\delta\gamma_2)(g_1,g_2,g_3)&=&L(g_1)\gamma_2(g_2,g_3)+
\gamma_2(g_1,g_2g_3)-\gamma_2(g_1g_2,g_3)-\gamma_2(g_1,g_2) \nn \eea

\medskip

As we will see below, the generalization of the coadjoint action and
its associated orbits naturally involves a cohomological structure. In order
to address this point, we consider the general elements above and choose
$A={\cal G}^*$ and $L=Coad$, i.e. the coadjoint action of $G$ on ${\cal
G}^*$. This choice leads in particular to the cohomology group
$H^1_{Coad}(G,{\cal G}^*)$, where a 1-cocycle
$\gamma:G\rightarrow {\cal G}^*$ is characterized by ($\delta\gamma\equiv
0$)
\bea \gamma(g'g)=Coad(g')\gamma(g)+\gamma(g') \label{cocsymp} \ \ ,
\eea meanwhile a 1-coboundary has the form ($\Delta_\mu\equiv\delta \mu$)
\bea \Delta_{\mu}=Coad(g)\mu-\mu \ \ , \ \ g\in G, \mu\in {\cal G}^* \
\ . \label{cobsymp} \eea
Symplectic cohomology $H_S(G,{\cal G}^*)$ is
defined out of this cohomology group by restricting the 1-cocycles to
functions $\gamma$ which satisfy the following antisymmetry condition on
its differential $\gamma^T$:
\bea \gamma^T(e)(X,Y)&\equiv&
\gamma^T(e)\cdot X(Y) \label{anti22}\\
\gamma^T(e)(X,Y)&=&-\gamma^T(e)(Y,X)\ \ \ \forall X,Y \in {\cal G} \ \ .
\nn \eea
The reason for this condition will be apparent in the next
subsection.

\medskip

\ni For the sake of completeness, we mention that the cohomology group
$H^2(G,U(1))$ we found in the previous section and which classifies the
central
extensions of the Lie group $G$, is obtained by setting
$A=U(1)$ and $L$ as the trivial representation in the general construction
above of Lie group cohomology.
Second and third
lines in (\ref{a2}) then define the expression of a coboundary and the
cocycle condition.

\subsection{Deformed coadjoint orbits}
As we have seen in Section 2.1, orbits of the coadjoint action of a group
$G$ on its coalgebra ${\cal G}^*$ constitute a class of symplectic
manifolds characterized in terms of group-theoretical structures.
Symplectic cohomology provides a way of introducing a notion of {\it
affine-deformations} of coadjoints actions which allow us to generalize
the notion of coadjoint orbit. Defining the mapping $g\mapsto Coad_\gamma
(g)$
\bea Coad_\gamma (g) \mu_0\equiv Coad(g)\mu_0+\gamma(g) \ \
\mu_0\in{\cal G}^* \label{accdef} \ \ , \eea
the condition for this
expression to actually define a (left) action of $G$ on ${\cal G}^*$, i.e.
$Coad_\gamma (g'g)\mu=Coad_\gamma (g')(Coad_\gamma (g)\mu)$, reduces to
expression (\ref{cocsymp}), which is simply the cocycle condition in
$H_S(G,{\cal G}^*)$.

\ni On the other hand, and denoting the orbit of $Coad_\gamma$ through the
point $\mu_0\in{\cal G}^*$ by $Orb_\gamma(\mu_0)$, we note that $\gamma$
functions which differ by a coboundary, (\ref{cobsymp}), define the same
set of orbits. In fact, since
\bea Coad_{\gamma+\Delta_{\mu}} (g)
\mu_0&=&Coad_{\gamma} (g) \mu_0 +Coad (g) \mu-\mu
\label{translation}\\
&=&Coad_\gamma(g) (\mu+\mu_0)-\mu \ \ \forall g\in G,
\forall \mu_0 \in {\cal G}^* \ \ , \nn \eea
we realize that
$Orb_{\gamma+\Delta_\mu}(\mu_0)$ and $Orb_\gamma(\mu_0+\mu)$ coincide
modulo a translation $\mu$. Therefore, if we allow $\mu$ to vary on ${\cal
G}^*$, each element in $H^1_{Coad}(G,{\cal G}^*)$ characterizes a family
of orbits (modulo translations) obtained from the deformed coadjoint
action on ${\cal G}^*$.

Finally, the antisymmetry condition on $\gamma^T(e)$ is necessary in order
to define a symplectic structure on $Orb_\gamma(\mu)$. If we define
\bea
\Gamma(X,Y)\equiv\gamma^T(e)\cdot X(Y)
\eea
the following theorem follows (\cite{So70}).

\medskip
\ni {\bf Theorem.} The orbit $Orb_\gamma(\mu)\subset {\cal G}^*$ admits a
symplectic form $\omega$ which is pointwise given by \bea
\omega_\nu(X,Y)=\nu([X_{\cal G},Y_{\cal G}])+\Gamma(X_{\cal G},Y_{\cal
G}), \ \ \nu\in Orb_\gamma(\mu), X,Y \in T_\nu Orb_\gamma(\mu) \ \ , \eea
where $\nu\in Orb_\gamma(\mu)$, $X,Y \in T_\mu Orb_\gamma(\mu)$ and
$X_{\cal G}\in {\cal G}$ is related to $X\in T_\mu Orb_\gamma(\mu)$ by
$X=coad_\gamma\; X_{\cal G} (\nu)$ and analogously for $Y$ and $Y_{\cal
G}$ (where $coad_\gamma\equiv (Coad_\gamma)^T(e)$).

\subsection{Convergence with the problem of central extensions}
\label{Convergence}

In Section 2 the techniques of GAQ were used in order to define a specific
symplectic structure that could be used as the support for the
Hamiltonian description of a classical system. The algorithm started from
a $U(1)$-centrally extended Lie group $\tilde{G}$, where the 2-cocycle
which defines the central extension permits the identification of the set
of variables building the sought phase space (for concreteness, those
coordinates associated with non-vertical vector fields which are absent
from the characteristic module of $\Theta\equiv {\theta^L}^{(\zeta)}$).
However, the object classifying the non-isomorphic $U(1)$-central
extensions of $G$, $H^2(G,U(1))$, is not fine enough in order to classify
the specific symplectic spaces, since some ambiguity still remains linked
to the choice of the particular coboundary for the 2-cocycle.

In an analogous manner, the approach followed in this Section, based on
deformed coadjoint actions, permits the classification of the different 
classes of deformed coadjoint orbits by the elements of $H_S(G,{\cal 
G}^*)$, but not the characterization of individual symplectic spaces. 

Therefore, the crucial mathematical structures of both approaches, 
$H^2(G,U(1))$ and $H_S(G,{\cal G}^*)$ respectively, need to be refined in 
order to account for such specific symplectic manifolds. 

Even at this intermediate step, a non-trivial convergence occurs between 
the conceptually different problems of classifying the central extensions 
of a given Lie group $G$ by $U(1)$, on the one hand, and the affine 
deformations of the coadjoint actions on ${\cal G}^*$, on the other hand.
In fact, the same object classifies the solutions to both problems, since 
$H^2(G,U(1))\approx H_S(G,{\cal G}^*)$. 

Although we shall dwell on this point in subsection 3.3.2., we can outline 
this equivalence by noting that, for simply connected groups, the 
isomorphism $H^2({\cal G},U(1))\approx H^2(G,U(1))$ 
\footnote{On behalf of concision, we avoid a presentation of
Lie algebra cohomology and refer the reader to standard references
like \cite{Jacobson}. We simply note that for simply connected groups,
Lie algebra cohomology emerges as an infinitesimal version of Lie
group cohomology.}
is satisfied and 
therefore it is enough to discuss the equivalence at the infinitesimal 
level. In fact, the cocycle condition (\ref{cocsymp}) implies the 
following condition on its differential $\Gamma(X,Y)$ 
\bea 
\Gamma([X,Y],Z)+\Gamma([Y,Z],X)+\Gamma([Z,X],Y)=0 \ \ ,\label{cocext}
\eea 
which, together with the antisymmetry condition (\ref{anti22}), 
$\Gamma(X,Y)=-\Gamma(Y,X)$ defines a 2-cocycle in $H^2({\cal G},U(1))$ 
(from the point of view of the central extensions of the Lie algebra 
${\cal G}$ (\ref{cocext}) is simply the Jacobi identity for the central 
generator in the Lie algebra; see \cite{Jacobson}). 
Likewise the infinitesimal expression of 
the coboundary condition (\ref{cobsymp}) implies, 
\bea 
\Gamma_{cob}(X,Y)=\mu([X,Y]) \ \ \hbox{ for some } \ \mu\in{\cal G}^* \ \ , 
\eea 
which is the coboudary condition in $H^2({\cal G},U(1))$.

\medskip

In section \ref{pseudocohomologia}, pseudo-extensions have been introduced 
as the element necessary to account for the specific symplectic manifolds, 
that we can construct out of a Lie group $G$ via a central extension of 
it. However the discussion was carried out only for the trivial class of 
$H^2(G,U(1))$. For the non-trivial cohomology classes the analysis was not 
so straightforward. However the convergence with the approach based on 
symplectic cohomology, and which aims directly at the problem of defining 
symplectic structures completely in terms of a Lie group, sheds a new 
light on the problem. From this perspective, the characterization of a 
specific symplectic structure for a (in general non-trivial) cohomology 
class $\gamma$ of $H^2({\cal G},U(1))\approx H^2(G,U(1))$, simply 
parallels the characterization of a particular orbit in the family of 
orbits defined by $Coad_\gamma$.

\subsubsection{Singularization of coadjoint orbits in symplectic cohomology}
\label{Singularization}

\ni In order to singularize a specific symplectic manifold out of the 
family defined by a cocycle in $H_S(G,{\cal G}^*)$, i.e. in order to 
characterize a particular deformed coadjoint orbit, we have two options: 
\begin{itemize}
\item[{\it i)}]We can fix a pair $(\gamma,\mu_0)$, where $\gamma$
specifies the cocycle which defines the deformation of the action and 
$\mu_0$ precises a point in the orbit. In this case, varying the second 
entry we scan all the possible orbits.
\item[{\it ii)}]Alternatively, we can fix the point $\mu_0$ in the coalgebra
and vary instead the representative of the cocycle $\gamma$ by modifying 
the coboundary, $\Delta_\mu$. Since the coalgebra is a linear space, there 
is a canonical choice for the fixed point $\mu_0$: the zero vector. We can 
see from expression (\ref{translation}) that the set of spaces constructed 
this way is the same that the one derived with option {\it i)}, 
although 
translated with respect to them in such a way that all these spaces share 
the zero vector in ${\cal G}^*$. 
\end{itemize}
However both characterizations are redundant since different pairs 
$(\gamma, \mu_0)$, or alternatively different specific representatives 
$\gamma+\Delta_\mu$, give rise essentially to the same orbits. Therefore 
it is necessary to establish an equivalence relationship in order to 
eliminate this ambiguity. The analysis of section 2.1. establishing the 
relationship between specific symplectic structures and pseudo-cohomology 
understood as a refinement of a true cohomology, suggest us to choose the 
characterization {\it ii)} for the deformed orbits. In fact, it directly 
leads to a refinement of symplectic cohomology, intrinsically tied to group 
cohomology. 

In this sense we have to determine under which conditions two coboundaries 
$\Delta_\mu$ and $\Delta_{\mu'}$ generate the same orbit. A direct 
computation shows that if there exists an element $h\in G$ such that 
$\mu=Coad_\gamma(h)\mu'$ (that is, if $\mu$ and $\mu'\in{\cal G}^*$ belong 
to the same $\gamma$-orbit) then 
\bea 
Coad_{\gamma+\Delta_\mu}(g)\,0=Coad_{\gamma+\Delta_{\mu'}}(gh)\,0
+\mu'-\mu \ \ \ \forall g\in G \ \ . \eea
Since $\mu'-\mu$ is independent 
of $g$, spaces spanned by the action of $Coad_{\gamma+\Delta_\mu}$ and 
$Coad_{\gamma+\Delta_\mu'}$ through the zero in ${\cal G}^*$ coincide, 
modulo a rigid translation. These orbits are trivially symplectomorphic, 
the symplectomorphism being this translation in ${\cal G}^*$. 

Summarizing with the language of symplectic cohomology, individual 
symplectic spaces associated with deformed coadjoints actions are classified 
by refinement of symplectic cohomology in such a way that two coboundaries 
$\Delta_\mu$ and $\Delta_{\mu'}$ are equivalent if $\mu$ and $\mu'$ belong 
to the same $\gamma-$orbit. \ni In other words, these individual symplectic 
spaces are classified by elements $\mu\in{\cal G}^*$ modulo the 
corresponding $\gamma$-deformed coadjoint action. Note the similarity with 
Definition 1, to which it directly generalizes in the context of deformed 
coadjoint orbits. 


\subsubsection{Pseudo-cohomology from symplectic cohomology}

In this section we see in a more systematic way the close relation between 
pseudo-cohomology and symplectic cohomology for the non-trivial classes 
($H^2(G,U(1))\neq 0$). The idea is to investigate how the coadjoint action
$Coad$ of $G$ on ${\cal G}^*$ is modified by a central extension. The 
result is that when $G$ is centrally extended by a 2-cocycle $\xi$, the 
coadjoint action of the extended group \Gt, denoted by $\widetilde{Coad}$, 
acting on $\widetilde{\calG}^*={\cal G}^*\times \mathbb{R}$, turns out to 
be 
\be
\widetilde{Coad}(\tilde{g})\tilde{\mu} =(Coad(g)\mu + \mu_\zeta F(g), 
\mu_\zeta) \label{Coadjuntaextendida} \ee 

\ni where $\tilde{g}=(g,\zeta)\in \Gtm,\,\zeta\in U(1)$ and 
$\tilde\mu=(\mu,\mu_\zeta)\in \widetilde{\calG}^*$. Here 
$F(g)\in{\calG}^*$, and it is related to the 2-cocyle $\xi$ through the 
quantization 1-form $\Theta$, by $F_i(g)=i_{\XR{i}}\Theta$. These 
functions are nothing other than the Noether invariants of the classical 
theory \cite{JMP23}. Observe that $\widetilde{Coad}(\tilde{g})$ does not
depend on $\zeta$, and that $\mu_\zeta$ does not change by this extended 
action (these two facts are related to the central character of $U(1)$). 
Since the case $\mu_\zeta=0$ reproduces the original coadjoint action 
$Coad$ of 
$G$, let us suppose $\mu_\zeta\neq 0$. 

{}From (\ref{Coadjuntaextendida}) it can be derived that 
$\widetilde{Coad}(g)$ can be restricted to the foliations of 
$\widetilde{\calG}^*$ of constant $\mu_\zeta$, which can be identified 
with ${\calG}^*$. Since $\widetilde{Coad}$ is an action, so 
it is its restriction, and this implies that $F(g)$ must verify the 
condition (this relation can also be checked by direct computation): 
\be
F(g'g)=Coad(g')F(g)+F(g') \ee 

Therefore Noether invariants are nothing other than 1-cocycles for the 
coadjoint action $Coad$ of $G$. Even more, they are symplectic, since its 
differential at the identity is precisely the Lie algebra 2-cocycle. 
Therefore, $\widetilde{Coad}$ can be identified with a deformed coadjoint 
action $Coad_{\gamma}$, with $\gamma(g)=\mu_\zeta F(g)$. 

Without losing generality, we can take $\mu_\zeta=1$. Let us see what 
happens to $Coad_{\gamma}$ when we add to $\xi$ a coboundary $\xi_\lambda$ 
generated by $\lambda(g)$. A simple calculation shows that $\gamma$ 
changes to $\gamma'=\gamma + \gamma_\lambda$, where $\gamma_\lambda$ is 
given by: 
\be
\gamma_\lambda(g) = Coad(g)\lambda^0 - \lambda^0 \ee 

Surprisingly, $\gamma_\lambda$ is a symplectic coboundary, associated with 
$\lambda^0\in {\cal G}^*$, and, what is more important, it depends just on 
$\lambda^0$, not on the particular choice of $\lambda$. This simple 
relation has deep consequences since it provides the close relation 
between pseudo-cocycles and symplectic cohomology. It also guides us in 
the correct definition of subclasses of pseudo-cocycles for the 
non-trivial case, using the characterization of single coadjoint orbits 
found in the symplectic cohomology setting (see Sec. \ref{Singularization}). 

According to this, and since the quantization 1-form $\Theta$ for any 
central extension \Gt characterized by the 2-cocycle $\xi$ always verifies
$\Theta|_e =(0,0,\ldots,1)$ (that is, $\mu=0$ and $\mu_\zeta=1$), we can 
singularize a deformed orbit in ${\cal G}^*$ by considering 
\be
\widetilde{Coad}(\tilde{g})\,\Theta|_e= (Coad_\gamma(g)\,0,1) = 
(Coad(g)\,0 + F(g),1) = (F(g),1) \ee 

\ni That is, this orbit is the image of the Noether invariants. This fact
simply affirms 
that Noether invariants parameterize classical phase spaces. 

The question now is that if we add to the 2-cocycle $\xi$ a pseudo-cocycle 
$\xi_\lambda$ generated by $\lambda(g)$ with gradient at the identity 
$\lambda^0$, does it define a new deformed coadjoint orbit? Can we define 
an equivalence relation among pseudo-cocycles as for the case of the 
trivial class?

Again, from the symplectic cohomology framework (see \ref{Singularization}), 
we have the answer. Firstly we define, 

\medskip

\noindent {\bf Definition 2:}\,\, {\it Two coboundaries $\xi_\lambda$ and 
$\xi_{\lambda'}$ with generating functions 
 $\lambda$ and $\lambda'$, respectively, define two cocycles 
$\xi+\xi_\lambda$, $\xi +\xi_{\lambda'}$ belonging to the 
same equivalence subclass $[\xi]$ of $[[\xi]]$ if and only if the gradients 
at the identity 
of the generating functions are related by:}
\be
\lambda^0{}'= Coad_\gamma(g)\lambda^0\,, \ee 

{\it \ni for some $g\in G$, where $Coad_\gamma$ stands for the deformed 
coadjoint action, which is equivalent to the coadjoint action 
$\widetilde{Coad}$ of \Gt on $\widetilde{\calG}^*={\cal G}^*\times 
\mathbb{R}$, where \Gt is the central extension associated with the two-cocycle $\xi$.}

\medskip

According to this definition, equivalent pseudo-extensions (for the 
non-trivial class $[[\xi]]$) are determined by generating functions whose 
gradient at the identity lie in the same coadjoint orbit of \Gt.

The ultimate justification of this definition is the following
proposition:

\medskip
\ni{\bf Proposition 3.} {\it Given a Lie group $G$ and a 2-cocycle
$\xi$ on $G$, consider the two coboundaries $\xi_{\lambda_1}$ and
$\xi_{\lambda_2}$ with generating functions $\lambda_1(g)$ and
$\lambda_2(g)$. Define the central extensions $\tilde{G}_1$ and $\tilde{G}_2$ characterized by the
two-cocycles $\xi+\xi_{\lambda_1}$ and $\xi+\xi_{\lambda_2}$, respectively, and construct the Quantization one-forms
$\Theta_1=\Theta+\Theta_{\lambda_1}$ and
$\Theta_2=\Theta+\Theta_{\lambda_2}$, following expressions (\ref{Theta})
and (\ref{Thetaprima}). The two symplectic spaces $\tilde{G}_1/
(G_{\Theta_1}\times U(1))$ and $\tilde{G}_2/ (G_{\Theta_2}\times U(1))$,
with symplectic forms $\omega_1$ and $\omega_2$, such that $d\Theta_1=\pi^*\omega_1$ and
$d\Theta_2=\pi^*\omega_2$, respectively, are symplectomorphic if there exists $h\in G$ such that:
\bea
\lambda_1^0=Coad_\gamma(h)\lambda_2^0,
\eea
the symplectomorphism being
given by $\widetilde{Ad}(\tilde{h})$:
\bea
d\Theta_1=(\widetilde{Ad}(\tilde{h}))^{*}d\Theta_2
\eea
\ni where $\tilde{h}$ is such that $p(\tilde{h})=h$, with
$p:\Gtm\rightarrow G$ the canonical projection.}

\ni {\it Proof:} Even though the result can be shown by direct calculation, 
the most 
straightforward derivation comes from splitting the central extension 
into two steps. 
Firstly, the central extension by $\xi$ alone is constructed, and this 
group $\tilde{G}$ is taken as the departing point for a second trivial 
extension by $\xi_{\lambda_1}$ and $\xi_{\lambda_2}$. The study of the 
trivial class in $\tilde{G}$ amounts for the study of that 
non-trivial class in $G$ characterized by the cocycle $\xi$.
At this point we can apply Proposition 1 
to the trivial extension of
$\tilde{G}$ and then take advantage of the identification 
between the non-trivial part (${\cal G}^*$-component in 
$\widetilde{\cal G}^*$) of $\widetilde{Coad}$ in $\tilde{G}$ and
$Coad_\gamma$ in $G$, which follows from expression (\ref{Coadjuntaextendida})
and its subsequent 
discussion. This leads directly to the claimed result.
 
\bigskip


As in the case of the trivial class, the correspondence between 
pseudo-cohomology classes in $[[\xi]]$ and "deformed" coadjoint orbits in 
${\cal G}^*$ is not onto. Only when we demand these coadjoint orbits to satisfy the integrality
condition (that is, to be quantizable), the correspondence with pseudo-cohomology classes is one-to-one.
The proof of this statement is the same as in the trivial case, see \cite{JMP23}.








\section{Final remarks}
In this paper we have established a neat characterization of the concept 
of pseudo-cohomology as the mathematical object classifying the single 
$G$-symplectic spaces that can be constructed out of a Lie group $G$. The role of 
symplectic cohomology has been crucial in this analysis: a) on the one 
hand it provides a clearer setting for the problem than the one based on 
central extensions; b) on the other hand, it offers a straightforward 
bridge for the translation of the results into the language of central 
extensions. 

This characterization is some-thing more than an academic problem, since these 
symplectic spaces constitute the classical phase spaces of the quantum 
theories associated with a fundamental symmetry. The {\it a priori} 
knowledge of the available classical structures provides a most valuable 
information in the study of the quantum theory. 
In this sense, and although this paper focuses on the discussion
of classical structures, a remark on their quantum counterparts is in
order. In fact, once the classification of symplectic spaces (deformed
coadjoint orbits) associated with a symmetry group has been done by means
of pseudo-cohomology, the question on the existence of non-equivalent  
quantizations corresponding to a given coadjoint orbit
$S$ naturally arises. As is well-known from Geometric Quantization
\cite{So70} (see also \cite{Woodhouse}) such a variety of non-isomorphic
quantum manifolds is classified by ${\pi_1}^*(S)$, i.e. the dual group
of the first homotopy group of the classical phase space. Although
this problem goes beyond the scope of the present work,
where we are interested in the classification of symplectic spaces, not in their quantization,
let us remark that when considering multiply connected coadjoint orbits
there exists the possibility of finding pseudo-cocycles associated with the same
coadjoint orbit and which
leads to non-equivalent representations (see the end of Sec. 2.1 for the
relation between pseudoextensions and quantization), hence to
non-equivalent quantizations.
These pseudo-cocycles are generated by non-homotopic functions
having the same gradient at the identity. An example of this situation
can be found in the case of the $SL(2,{\mathbb R})$ group which
admits two non-equivalent classes of unirreps associated with
the multiply connected coadjoint orbits \cite{Ba47}. A precise analysis of
this example can be seen in \cite{AG03}. As a consequence,
the classification of non-equivalent representations associated with the
same coadjoint orbit would require a further refinement in the characterization of 
pseudo-cohomology classes.

It should be stressed that although pseudo-cohomology with values on $U(1)$ classifies 
quantizable $G$-symplectic manifolds through the integrality condition, general 
classical $G$-symplectic manifolds can be regained by considering pseudo-cohomology with 
values on the additive group $R$, rather than $U(1)$. In fact, the Group Approach to 
Quantization recovers Classical Mechanics, in the Hamilton-Jacobi version, by just 
considering the additive group ${\mathbb R}$ instead of the multiplicative one 
$U(1)$, the former being a local approximation to the latter.

Finally, and coming back to the original motivation for pseudo-cohomology in terms 
of In\"on\"u-Wigner contractions, let us remark that, conversely, given a 
pseudo-cohomology class and the corresponding quantization group $\tilde{G}$, with 
quantization form $\Theta$, an In\"on\"u-Wigner contraction with respect the characteristic 
subgroup $G_\Theta$ of $\Theta$ automatically leads to a contracted group $\tilde{G}_c$
which proves to be a non-trivial extension by $U(1)$ of the contraction $G_c$ of $G$ by the same 
subgroup $G_\Theta$.

\end{document}